\documentclass{article}

\usepackage{PRIMEarxiv}

\usepackage[utf8]{inputenc} % allow utf-8 input
\usepackage[T1]{fontenc}    % use 8-bit T1 fonts
\usepackage{hyperref}       % hyperlinks
\usepackage{url}            % simple URL typesetting
\usepackage{booktabs}       % professional-quality tables
\usepackage{amsfonts}       % blackboard math symbols
\usepackage{nicefrac}       % compact symbols for 1/2, etc.
\usepackage{microtype}      % microtypography
\usepackage{lipsum}
\usepackage{multicol}
\usepackage{fancyhdr}       % header
\usepackage{caption}        % caption
\usepackage{graphicx}       % graphics
\graphicspath{{media/}}     % organize your images and other figures under media/ folder
\usepackage{authblk}
\usepackage{cleveref}

\title{\textbf{A SMART FRIDGE WITH AI-ENABLED FOOD COMPUTING}
}

\usepackage{authblk}

\author[1,2,3,*]{\textbf{Khue Nong Thuc\_Computer Science}}
\author[1,2,3]{\textbf{Khoa Tran Nguyen Anh\_Computer Science}}
\author[1,2,3,*]{\textbf{Tai Nguyen Huy\_Computer Engineering}}
\author[1,2,3]{\textbf{Du Nguyen Hao Hong\_Computer Science}}
\author[1,2,3]{\textbf{Khanh Dinh Ba\_Computer Science}}

\affil[1]{Faculty of Computer Science and Engineering, Ho Chi Minh City University of Technology (HCMUT), 268 Ly Thuong Kiet Street, District 10, Ho Chi Minh City, Vietnam}
\affil[2]{Office for International Study Programs, Ho Chi Minh City University of Technology (HCMUT), 268 Ly Thuong Kiet Street, District 10, Ho Chi Minh City, Vietnam}
\affil[3]{Vietnam National University Ho Chi Minh City, Linh Trung Ward, Thu Duc City, Ho Chi Minh City, Vietnam}
\affil[*]{Corresponding author: \href{mailto:khue.nongthuc@hcmut.edu.vn}{khue.nongthuc@hcmut.edu.vn} and \href{mailto: tai.nguyenraphael@hcmut.edu.vn}{tai.nguyenraphael@hcmut.edu.vn}}

\begin{document}
% formate figure
% formate figure
\captionsetup[figure]{labelfont={bf,it}}

\maketitle

\begin{abstract}
The Internet of Things (IoT) plays a crucial role in enabling seamless connectivity and intelligent home automation, particularly in food management. By integrating IoT with computer vision, the smart fridge employs an ESP32-CAM to establish a monitoring subsystem that enhances food management efficiency through real-time food detection, inventory tracking, and temperature monitoring. This benefits waste reduction, grocery planning improvement, and household consumption optimization. In high-density inventory conditions, capturing partial or layered images complicates object detection, as overlapping items and occluded views hinder accurate identification and counting. Besides, varied angles and obscured details in multi-layered setups reduce algorithm reliability, often resulting in miscounts or misclassifications. Our proposed system is structured into three core modules: data pre-processing, object detection and management, and a web-based visualization. To address the challenge of poor model calibration caused by overconfident predictions, we implement a variant of focal loss that mitigates over-confidence and under-confidence in multi-category classification. This approach incorporates adaptive, class-wise error calibration via temperature scaling and evaluates the distribution of predicted probabilities across methods. Our results demonstrate that robust functional calibration significantly improves detection reliability under varying lighting conditions and scalability challenges. Further analysis demonstrates a practical, user-focused approach to modern food management, advancing sustainable living goals through reduced waste and more informed consumption.

\end{abstract}

% keywords can be removed

\twocolumn

\section{Introduction}
Food waste has emerged as an issue with significant environmental, economic, and social implications. According to the Food and Agriculture Organization (FAO) of the United Nations, approximately one-third of all food produced globally—about 1.3 billion tons—is wasted each year \cite{FoodandAgriculture}. A considerable portion of this waste occurs at the household level, often due to poor visibility and management of perishable items such as vegetables. Furthermore, improper storage conditions, particularly inconsistent temperature and humidity in refrigerators, accelerate food spoilage, leading to unnecessary disposal of otherwise edible food. In this context, there is an increasing demand for systems that assist users in monitoring food inventory and maintaining optimal storage conditions. The Smart Fridge system is proposed as an IoT-based solution that enables monitoring of food stored in refrigerators.

This system utilizes an ESP32-CAM module to capture images of the fridge’s interior, with a particular focus on vegetables. These images are processed using the YOLO object detection model, which offers high accuracy in detecting and classifying different types of vegetables under various environmental conditions. In addition to visual recognition, the system incorporates temperature and humidity sensors to monitor the internal environment of the refrigerator, allowing users not only to track the types and quantities of stored vegetables but also to ensure that they are preserved under suitable conditions. The captured data is processed and presented through a web-based interface, offering users an intuitive and accessible platform to monitor fridge contents remotely. \cite{nguyen2020} This interface can be enhanced with alert systems, notifying users of low inventory levels, suboptimal storage conditions, or impending spoilage.

In this work, the system is developed with three primary objectives:
 \begin{itemize}
     \item The input of our system is color bits, static images of vegetables in a fridge captured by ESP32-CAM.
     \item The inside-fridge pictures will be updated per minute, and the vegetables in the images will be detected by using the YOLO model.
     \item The expected output is static pictures with labels, a static table with the quantity of each kind of current vegetables in the fridge, and a visualization of the temperature and humidity. 
 \end{itemize}

\section{Proposed system design}
\subsection{System design}
The Smart Fridge AI system consists of three main modules: data processing \& object detection, data subscription \& storage, and web-based visualization. The ESP32-CAM captures images of the fridge interior, which are processed by a YOLO model on the YOLO UNO board to detect and classify fruits and vegetables. This board also collects temperature data and transmits both image and sensor information to the CoreIoT cloud via MQTT. The backend, developed using NodeJS, retrieves and stores this data in a MongoDB database, enabling efficient historical tracking and analysis. The frontend, built with ReactJS, presents real-time annotated images, a summary of detected items, and environmental trends.

\begin{figure}[h!]
    \centering
    \includegraphics[width=0.45\textwidth]{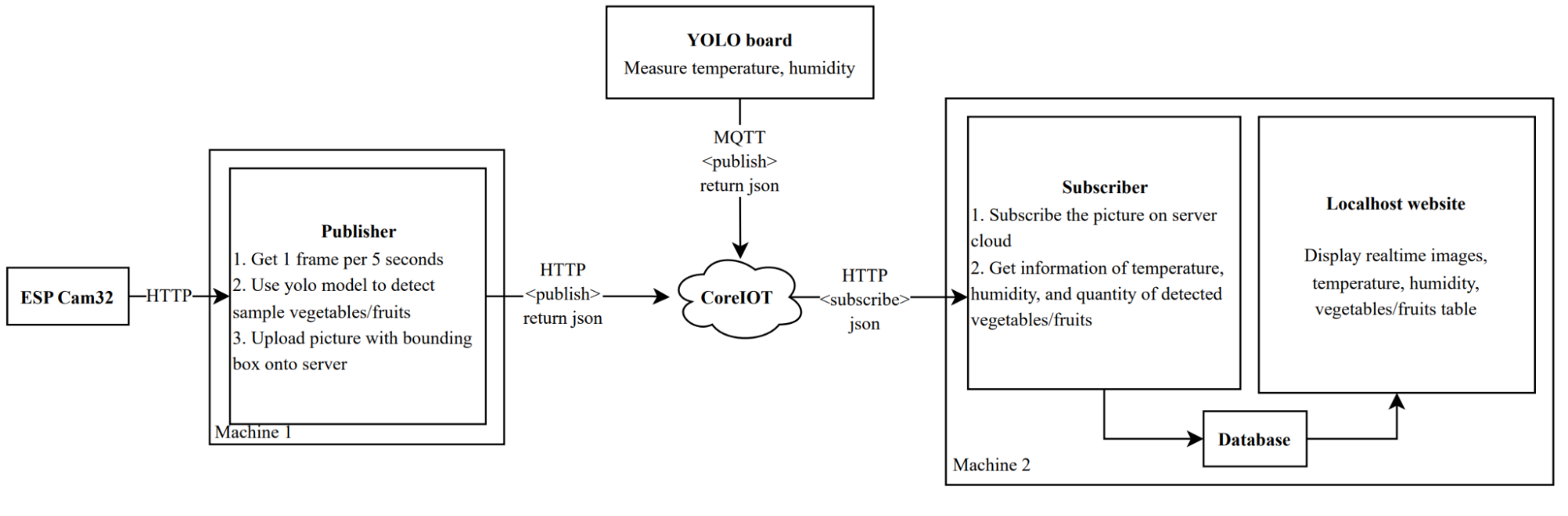}
    \caption{System design diagram}
    \label{fig:systemdesign}
\end{figure}

To enable efficient communication between distributed components, the system adopts a Publish–Subscribe architecture using the MQTT protocol. This approach decouples components by allowing them to communicate asynchronously via a central broker, enhancing scalability and flexibility. In this setup, the YOLO Uno module acts as a publisher, analyzing image data and detecting predefined behaviors such as falls, safety violations, or intrusions. When such events are detected, it sends JSON-formatted alert messages to specific MQTT topics, including metadata like event type, timestamp, and object count.

On the subscriber side, the Subscriber call back module listens to these topics using an MQTT client. Upon receiving a message, it parses the alert and triggers appropriate actions, typically sending an HTTP POST request to a physical device (e.g., alarm or warning system) to respond in real time. Additionally, alerts can be logged into a structured database for historical analysis. This architecture ensures low-latency communication, real-time responsiveness, and a clean separation between detection and action, making it ideal for smart, event-driven IoT applications.

\subsection{Data Processing \& Object Detection Module}
\begin{figure}[h!]
    \centering
    \includegraphics[width=0.45\textwidth]{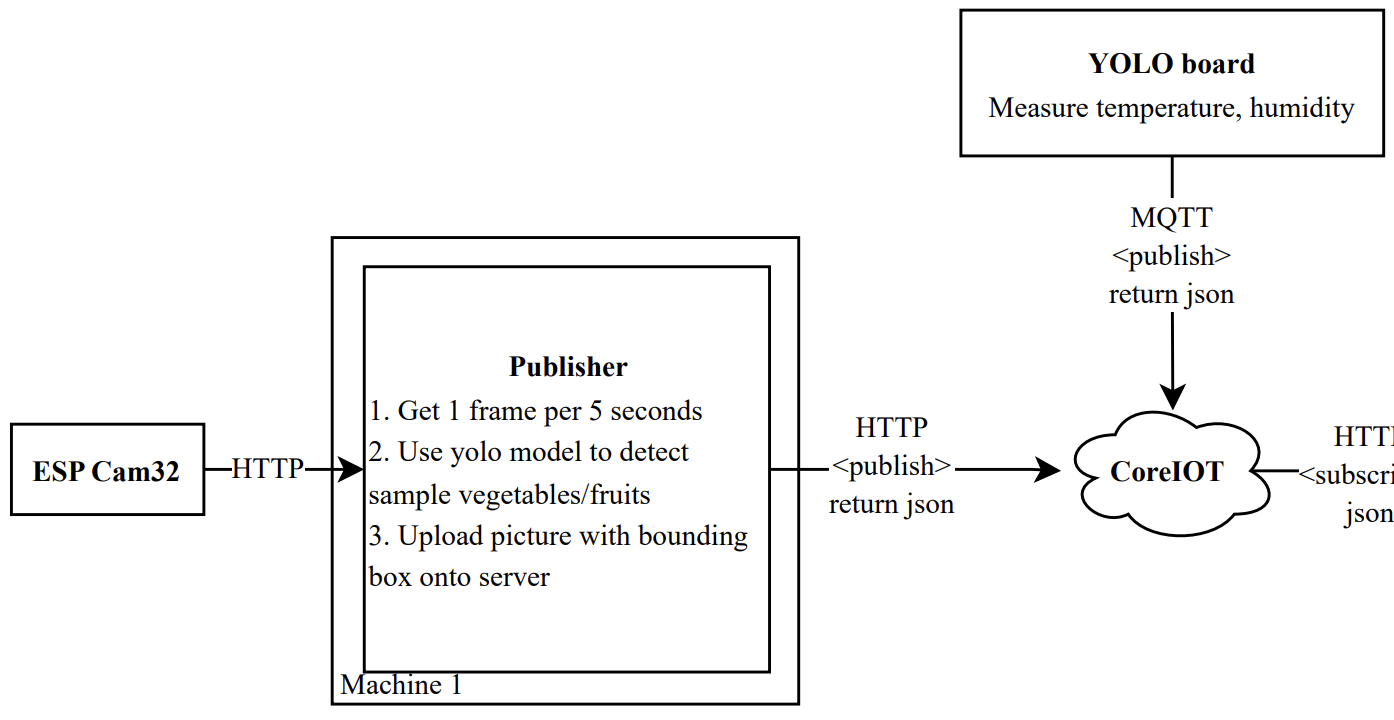}
    \caption{Design diagram of module 1}
    \label{fig:module1}
\end{figure}

The first module is responsible for capturing images, detecting objects, and publishing the processed data to the cloud for further analysis. The ESP32-CAM is used to capture images inside the fridge at regular intervals. These images are then processed by a YOLO-based object detection model to identify and classify vegetables and fruits. After detection, bounding boxes are applied to the images, and the processed data is published to the CoreIoT platform using the HTTP protocol, as show in \textbf{\autoref{fig:module1}}.

The YOLO UNO board plays a crucial role in this module, as it is responsible for performing object detection efficiently. In addition to detecting objects, the YOLO UNO board also collects temperature data from sensors placed inside the fridge. This temperature data, along with the detected object counts, is formatted into a structured JSON message and sent to the CoreIoT cloud.
The MQTT protocol is used for transmission between the YOLO board and CoreIOT, ensuring lightweight and efficient data transfer. The processed images, object detection results, and temperature data are published as JSON messages, making them accessible to other modules in the system. By implementing this module, the system ensures real-time monitoring of fridge contents, providing valuable insights for food management and waste reduction.

\subsection{Data Subscription \& Storage Module}
The second module is responsible for subscribing to the data published by the first module, processing it, and storing it in a database. This module acts as the backend of the system, retrieving images and JSON messages from the CoreIoT platform via HTTP api subscription. The retrieved data includes labeled images with bounding boxes, temperature readings, and object count information.

\begin{figure}[h!]
    \centering
    \includegraphics[width=0.45\textwidth]{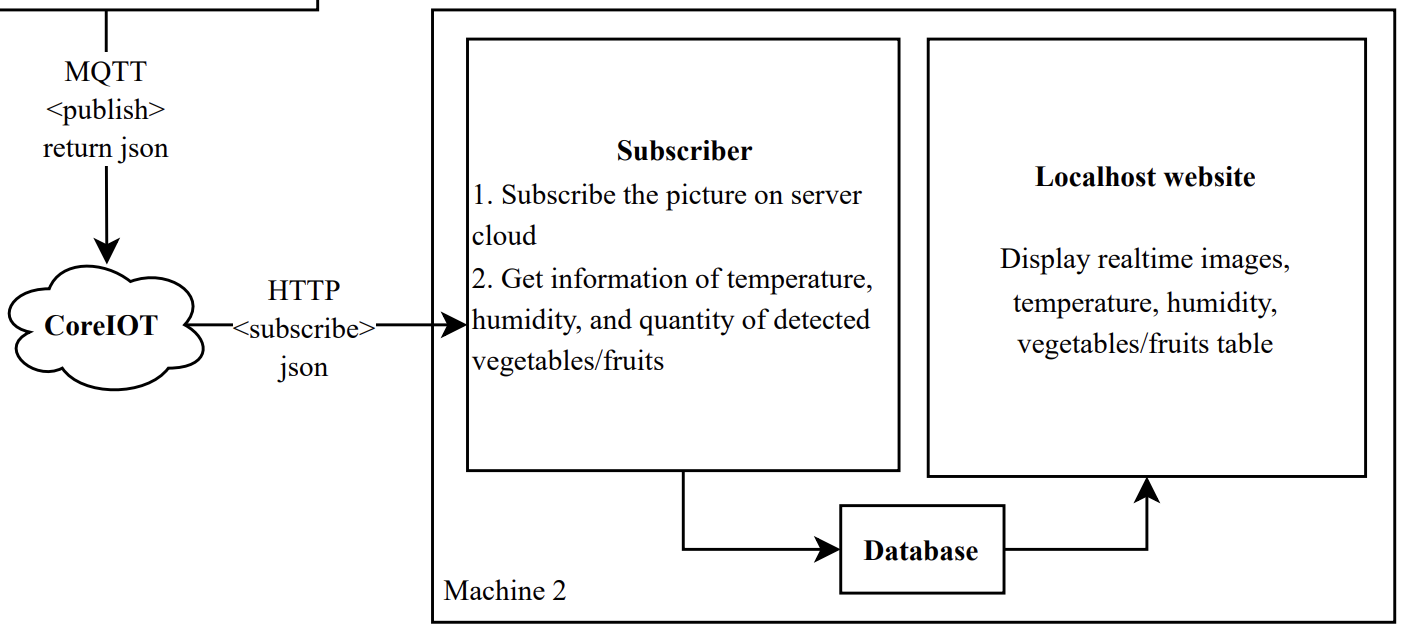}
    \caption{Design diagram of module 2}
    \label{fig:module2}
\end{figure}

Once the data is received, it is parsed and stored in a structured format within a database. The database is designed to efficiently store historical fridge data, enabling users to track changes over time. The system ensures that each received image is correctly linked with its corresponding object detection results and temperature values.

This module is essential for maintaining an organized and reliable dataset that can be accessed later for analytics, trend detection, and user notifications. By storing the data in a database, the system provides long-term tracking capabilities, helping users understand their consumption habits and detect patterns in fridge usage. 

\subsection{Web-based Visualization Module}
The third module focuses on presenting the collected data to the user through a web-based interface. The web application is hosted on a local machine and serves as the primary interaction point for users. It retrieves data from the database and displays it in an intuitive and user-friendly manner.

\begin{figure}[h!]
    \centering
    \includegraphics[width=0.45\textwidth]{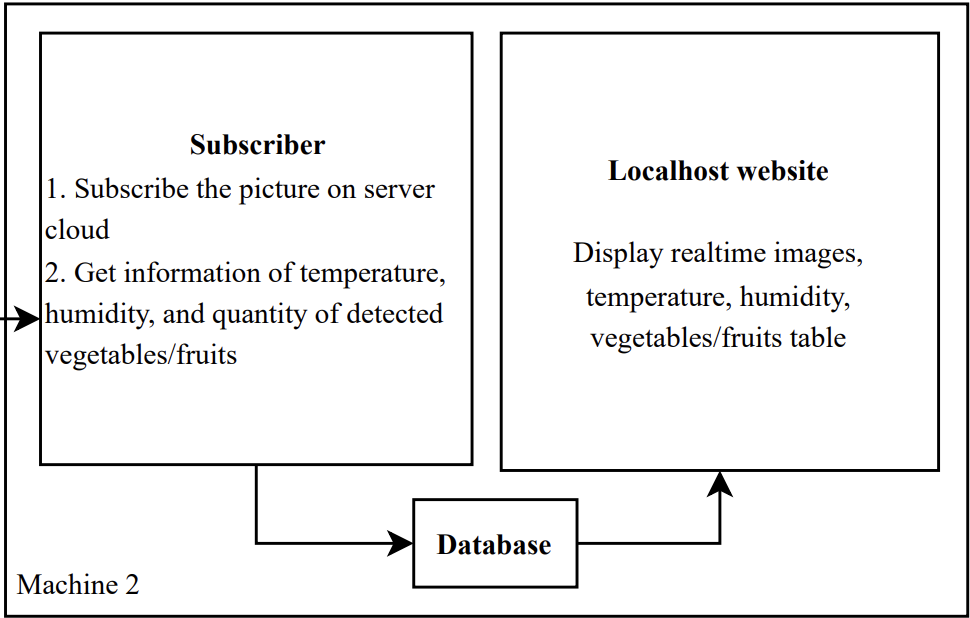}
    \caption{Design diagram of module 3}
    \label{fig:module3}
\end{figure}

The interface consists of multiple components, including a real-time display of the latest fridge images, a table summarizing the detected vegetables and their quantities, and a temperature and humidity visualization graph. The images displayed on the web interface include bounding boxes, making it easy for users to identify the detected objects. The table provides a structured view of available fridge contents, allowing users to monitor their food inventory efficiently. In addition to displaying current data, the web application allows users to access historical records and analyze trends over time. By integrating real-time updates and historical tracking, this module ensures that users have complete visibility into their fridge’s status, helping them make informed decisions regarding food storage and consumption.

\begin{figure}[h!]
    \centering
    \includegraphics[width=0.45\textwidth]{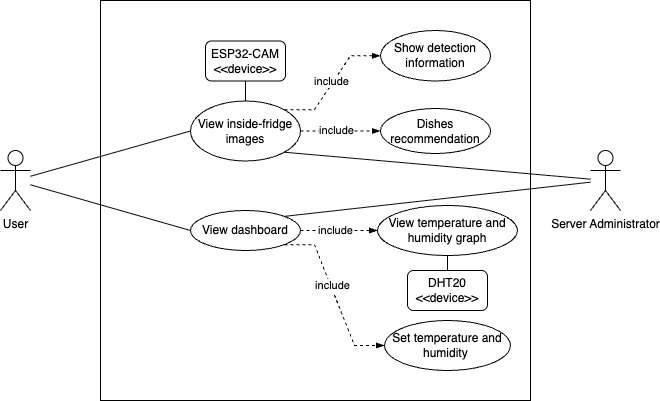}
    \caption{Use case diagram of the whole system}
    \label{fig:usecase}
\end{figure}

\textbf{Database}

MongoDB stores data in three collections:
\begin{itemize}
    \item images: stores timestamps and image data.
    \item counts: stores object list with name and quantity.
    \item fridgestats: stores temperature and humidity records.
    \item users: stores accounts and passwords of users.
\end{itemize}

\textbf{Back-end}

The backend is implemented using NodeJS and ExpressJS. It includes:
\begin{itemize}
    \item Routes for receiving data from CoreIoT.
    \item APIs for the frontend to fetch the latest readings.
    \item Post endpoints to update temperature/humidity settings.
    \item CoreIoT adapter module to communicate with the external IoT API.
\end{itemize}

\textbf{Front-end}

The front-end uses:
\begin{itemize}
    \item Axios for API calls.
    \item setInterval or useEffect with timers for auto-refresh.
    \item Rechart.js library for visual representation.
    \item Image components to show fridge state.
\end{itemize}

\section{Related work in loss formulation}
\subsection{Focal Loss}
This model was trained with a calibration-aware version of Focal Loss to treat the over-confident predictions. Although Focal Loss is designed to handle class imbalance, this can lead the model to be excessively confident in its outputs, even when incorrect. This behaviour can degrade the reliability of the model in applications in which probability calibration is essential.

\begin{equation}
\mathcal{L}_{FL(x,y)} = - \sum_{i=1}^{K} y_i \,(1 - q_i(x))^{\gamma} \log(q_i(x)) \; \cite{tao2023dual}
\end{equation}
with $\gamma$ representing a hyperparameter known as the focusing parameter, 
$y_i$ is the true label, and $q_i(x)$ is the prediction probability 
of the model for the class $i$.

\subsection{Adaptive Focal Loss}
Based on the standard version of Focal Loss, the adaptive version dynamically adjusts the focusing parameter to better align with the difficulties in learning across classes. Nevertheless, despite the nature of adaptive, this model still produced poorly calibrated outputs, with a similar or stronger trend toward over-confidence. This suggests that whereas adaptivity may help balance learning across the distribution in data, it does not necessarily result in better-calibrated probability estimates.

\begin{equation}
    \gamma_{t,b} = \gamma_{t-1,b} * exp(\lambda(C_{val,b} - A_{val,b})) \; \cite{ghosh2023adafocal}
\end{equation}

\subsection{Binary Cross Entropy (BCE)}
In contrast, the model trained with Binary Cross Entropy \cite{bce} demonstrated significantly better calibration. It produced more conservative probability estimates that were better aligned with the actual correctness of predictions. This conservative behavior indicates that BCE, while not explicitly designed to handle class imbalance, avoids the overconfidence issues introduced by Focal Loss variants.

\begin{figure}[h!]
    \centering
    \includegraphics[width=0.45\textwidth]{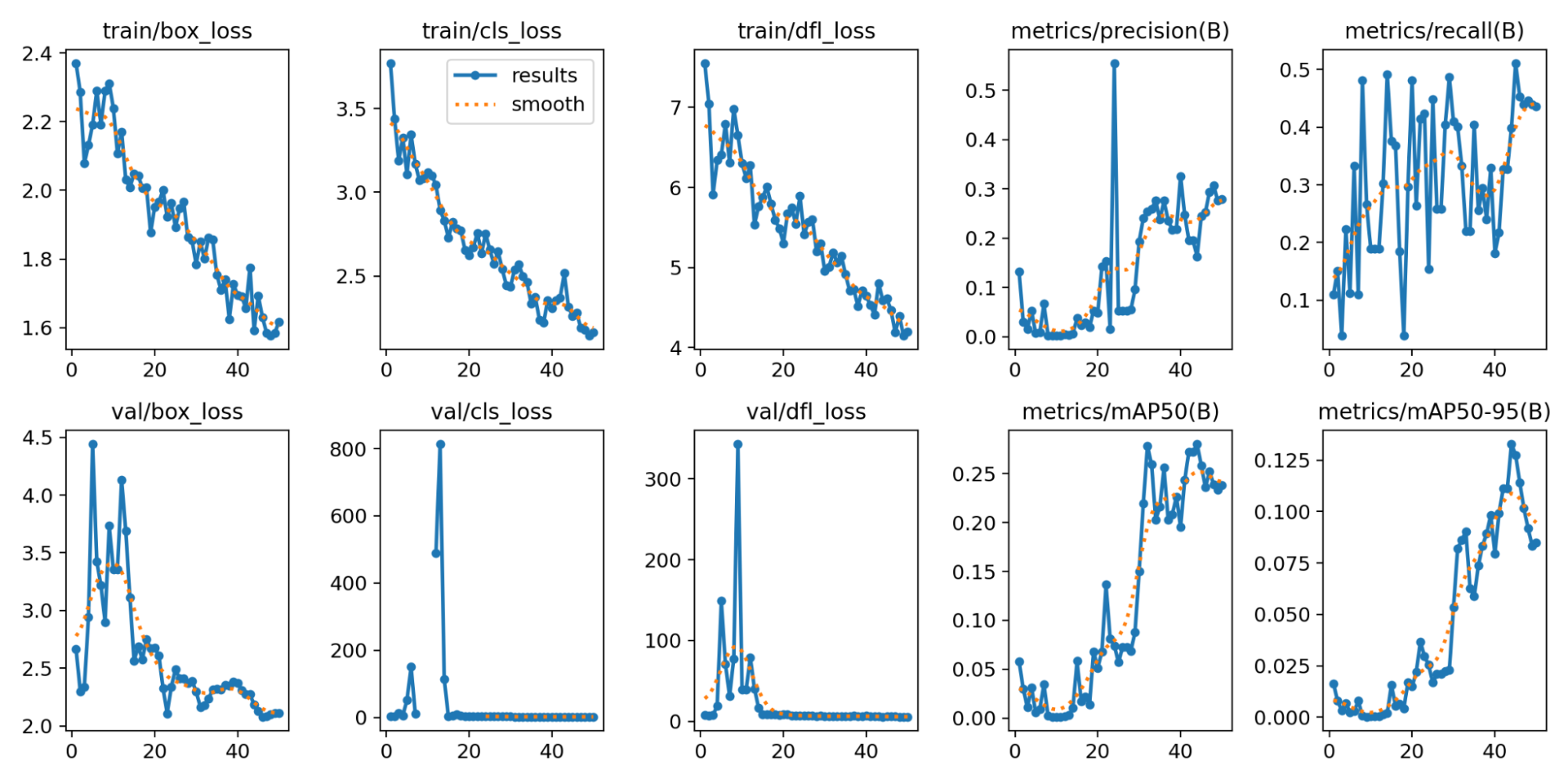}
    \caption{Before applying focal loss (epoch = 50)}
    \label{fig:beforefocalloss}
\end{figure}

\begin{figure}[h!]
    \centering
    \includegraphics[width=0.45\textwidth]{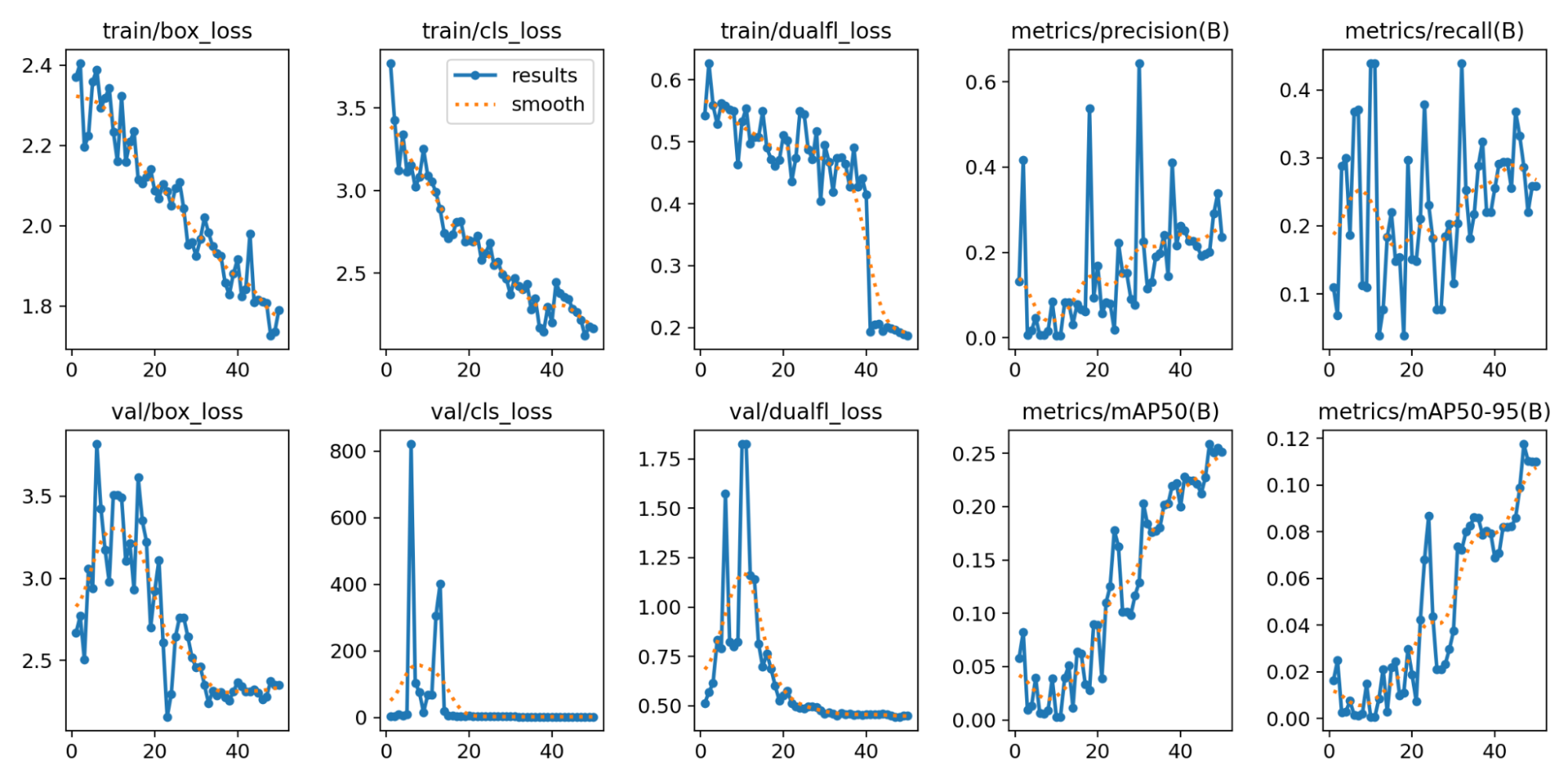}
    \caption{After applying focal loss (epoch = 50)}
    \label{fig:afterfocalloss}
\end{figure}

\section{Numerical results and discussion}
\subsection{Testbed setup}
The detection feature is implemented using Python, PyTorch, and the Ultralytics YOLO model \cite{wang2022yolov7}. The dataset utilized for training and testing is a combination of real images captured by the ESP32-CAM and additional images sourced from online repositories. These images are pre-processed using Roboflow, a tool designed to streamline dataset preparation and augment image data for enhanced model performance. This approach ensures that the detection model can effectively identify and classify objects, particularly vegetables and fruits, under varying environmental conditions.

In order to improve the performance and efficiency, we apply some loss functions to reduce the loss of the YOLO model. There are focal loss, adaptive focal loss, and BCE.

\subsection{Performance results}
We attempted to train the YOLO model on the training set and validation set with 50 epochs and visualized the results. Each set of subplots shows metrics including box loss, classification loss, DFL/dualFL loss, precision, recall, and mean Average Precision (mAP). 

Before applying focal loss to train the YOLO model, the loss values in the training set decrease gradually. However, in the validation set, the loss values increase dramatically in the early epochs and fluctuate heavily after that. The mAP50 increases unstably and inconsistently. Therefore, we apply the focal loss to the model in order to mitigate the overfitting.

After integrating focal loss into the YOLO model, the training losses continue to decrease consistently, while the validation losses become significantly more stable and lower in magnitude compared to the baseline. The mAP50 metric shows a clear upward trend with reduced fluctuations, reflecting enhanced detection accuracy and consistency. These improvements confirm that the application of focal loss successfully mitigates overfitting and improves the model’s ability to perform reliably on unseen data.

\begin{figure}[h!]
    \centering
    \includegraphics[width=0.45\textwidth]{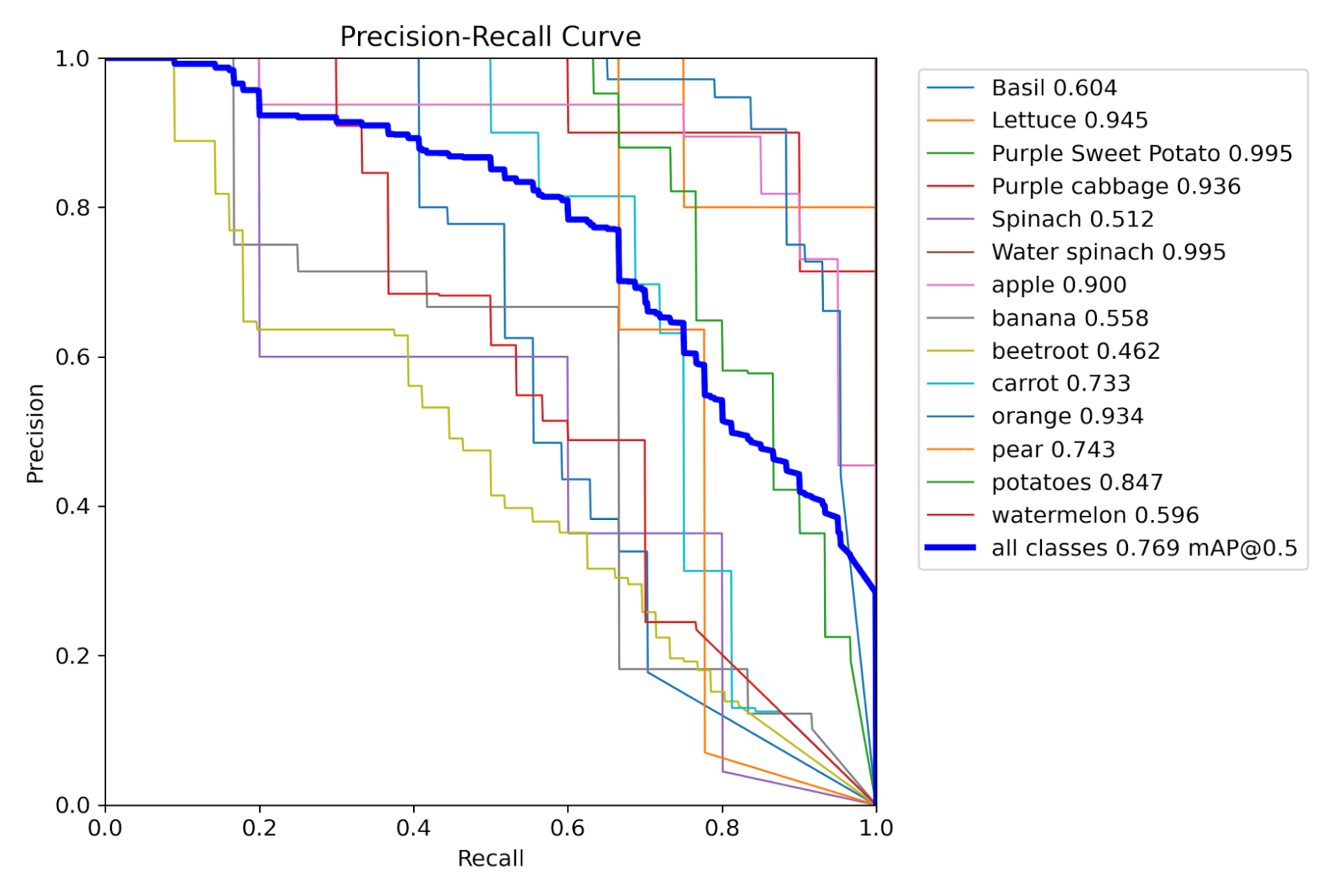}
    \caption{Precision-recall curve of the model after applying focal loss}
    \label{fig:pre_re}
\end{figure}

\begin{figure}[h!]
    \centering
    \includegraphics[width=0.45\textwidth]{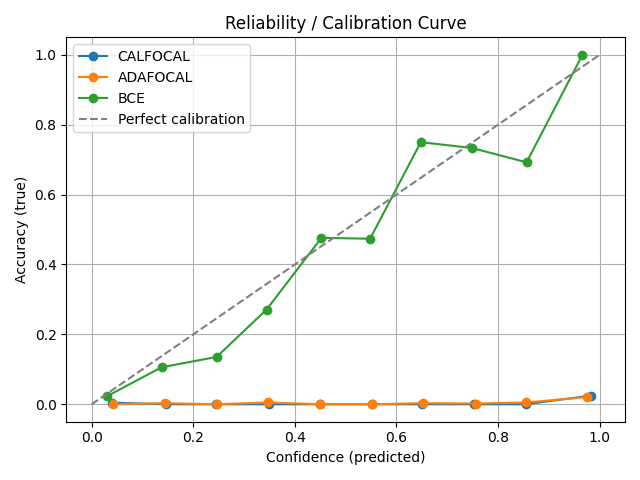}
    \caption{Reliability/Calibration curve of 3 models applied focal loss, adaptive focal loss, and BCE}
    \label{fig:re_ca}
\end{figure}

The precision-recall curve shows that most classes achieve high detection performance after applying focal loss, with several (e.g., Purple Sweet Potato, Water Spinach, Apple) reaching over 0.450 average precision. Some classes, such as Beetroot and Spinach, perform less consistently. The overall mAP@0.5 across all classes reaches \textbf{0.769}, indicating improved accuracy and robustness of the model, as show in \textbf{\autoref{fig:re_ca}}.

The calibration curve compares the alignment between predicted confidence and actual accuracy across three loss functions. The BCE model shows relatively good calibration, with predictions following the diagonal line of perfect calibration. In contrast, models using focal loss (CALFOCAL) and adaptive focal loss (ADAFOCAL) demonstrate severe underconfidence, with predicted confidence values not corresponding to actual accuracy. This indicates poor model calibration, suggesting that while focal-based losses may improve classification performance, they may require additional techniques for better confidence alignment.

\begin{figure}[h!]
    \centering
    \includegraphics[width=0.45\textwidth]{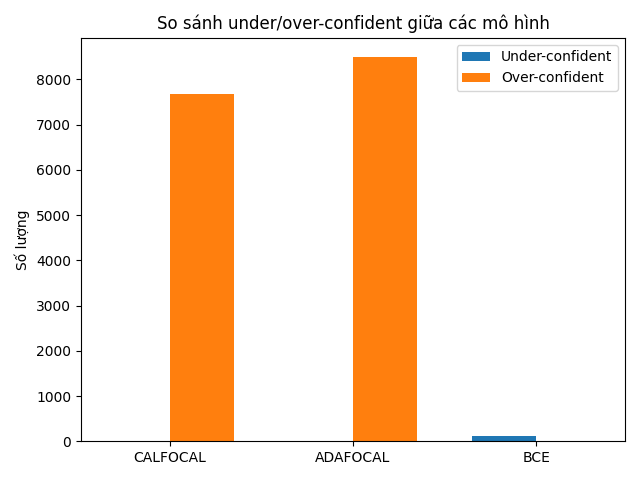}
    \caption{Comparison of under-confidence and over-confidence between models}
    \label{fig:comparison}
\end{figure}

\begin{figure}[h!]
    \centering
    \includegraphics[width=0.45\textwidth]{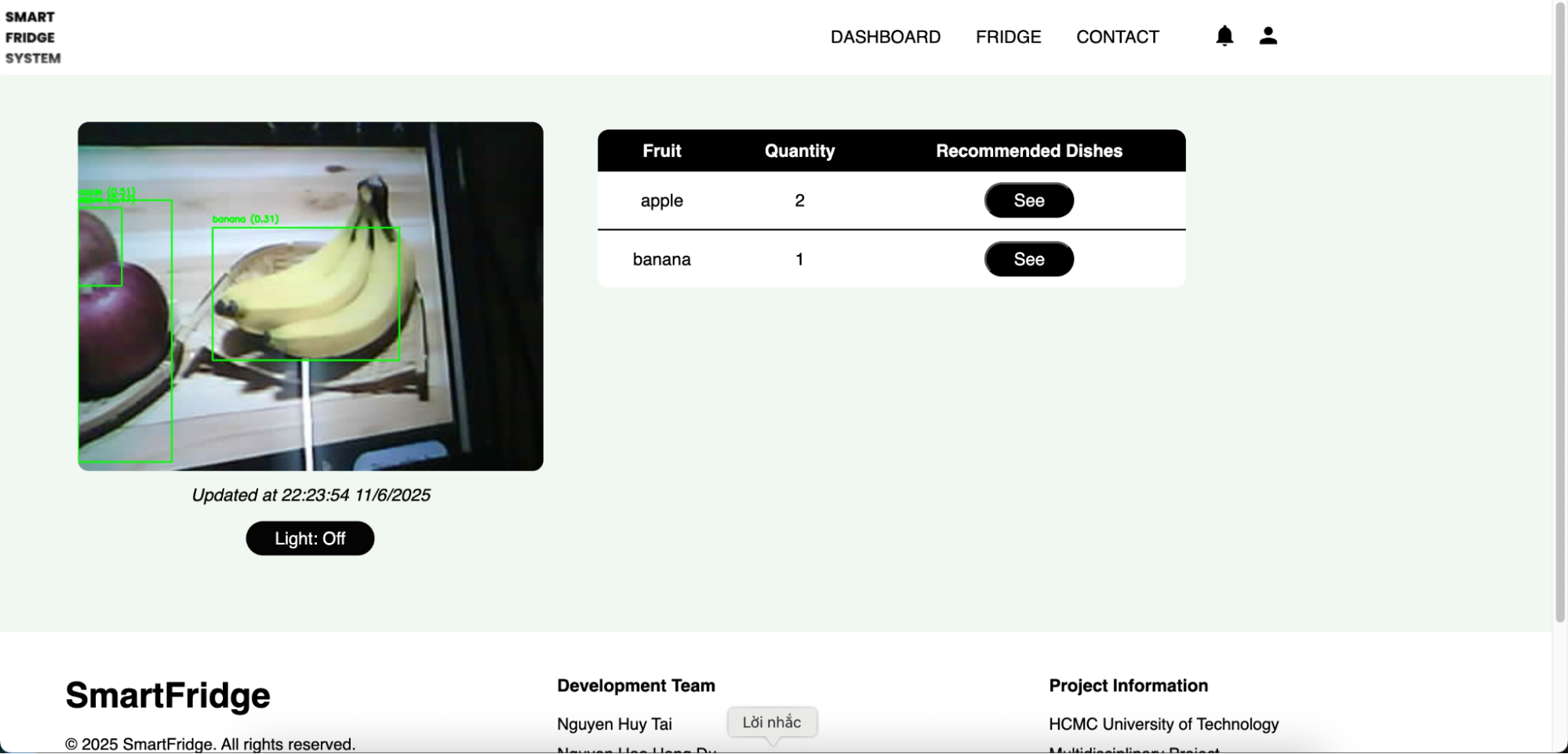}
    \caption{The website page displays the real-time object detection with ESP32-CAM}
    \label{fig:website}
\end{figure}

\begin{figure}[h!]
    \centering
    \includegraphics[width=0.45\textwidth]{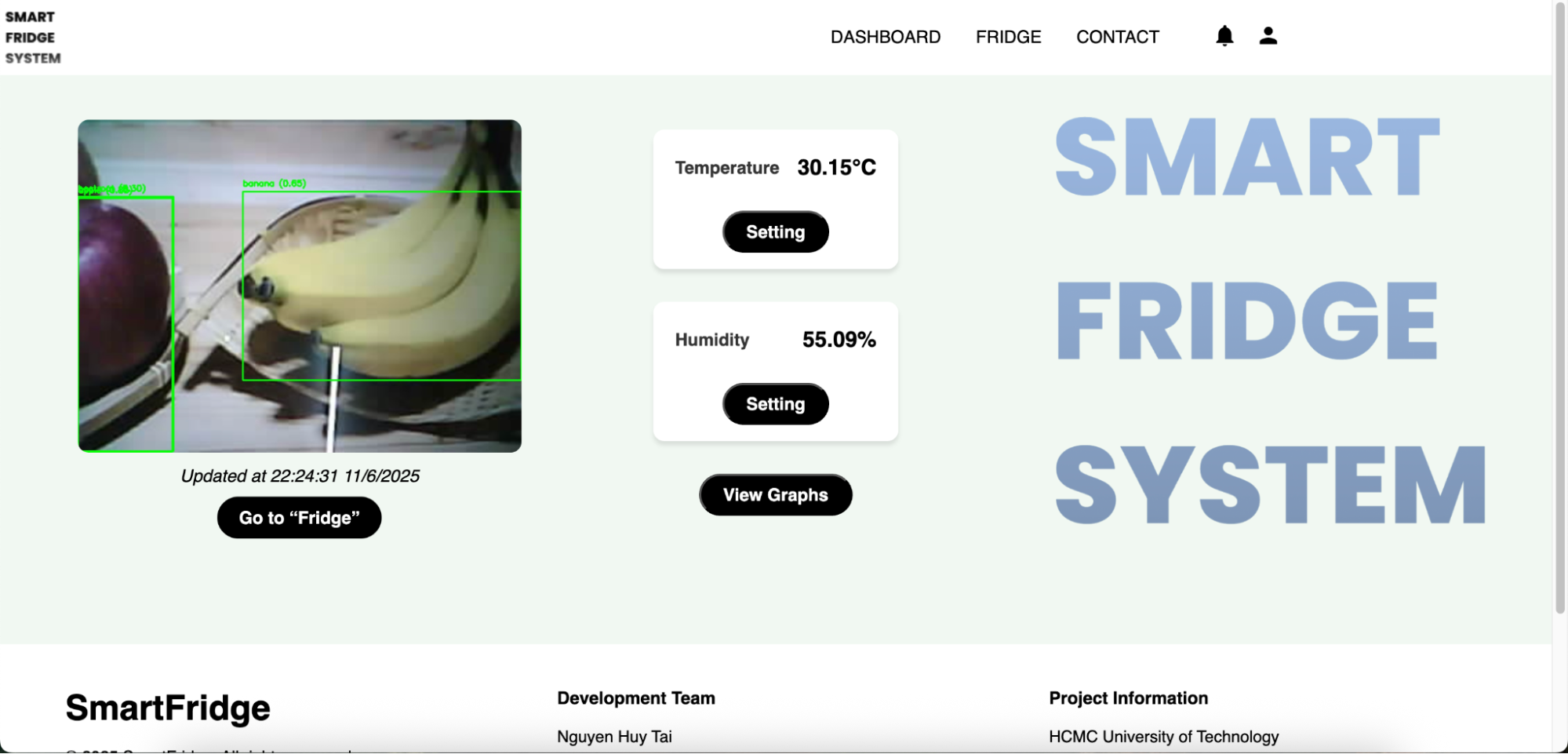}
    \caption{Home page after signing in successfully}
    \label{fig:homepage}
\end{figure}

We connect the ESP32-CAM to the system to perform end-to-end testing. The camera captures real-time images of the refrigerator’s interior, which are then processed by the YOLO detection model. The detection results, including annotated images and object quantities, are displayed on the web interface. As shown in {\bfseries \autoref{fig:website} and \autoref{fig:homepage}}, the system identifies and labels the fruits detected in the frame, while also generating a table of their quantities along with suggested recipes.

\begin{figure}[h!]
    \centering
    \includegraphics[width=0.45\textwidth]{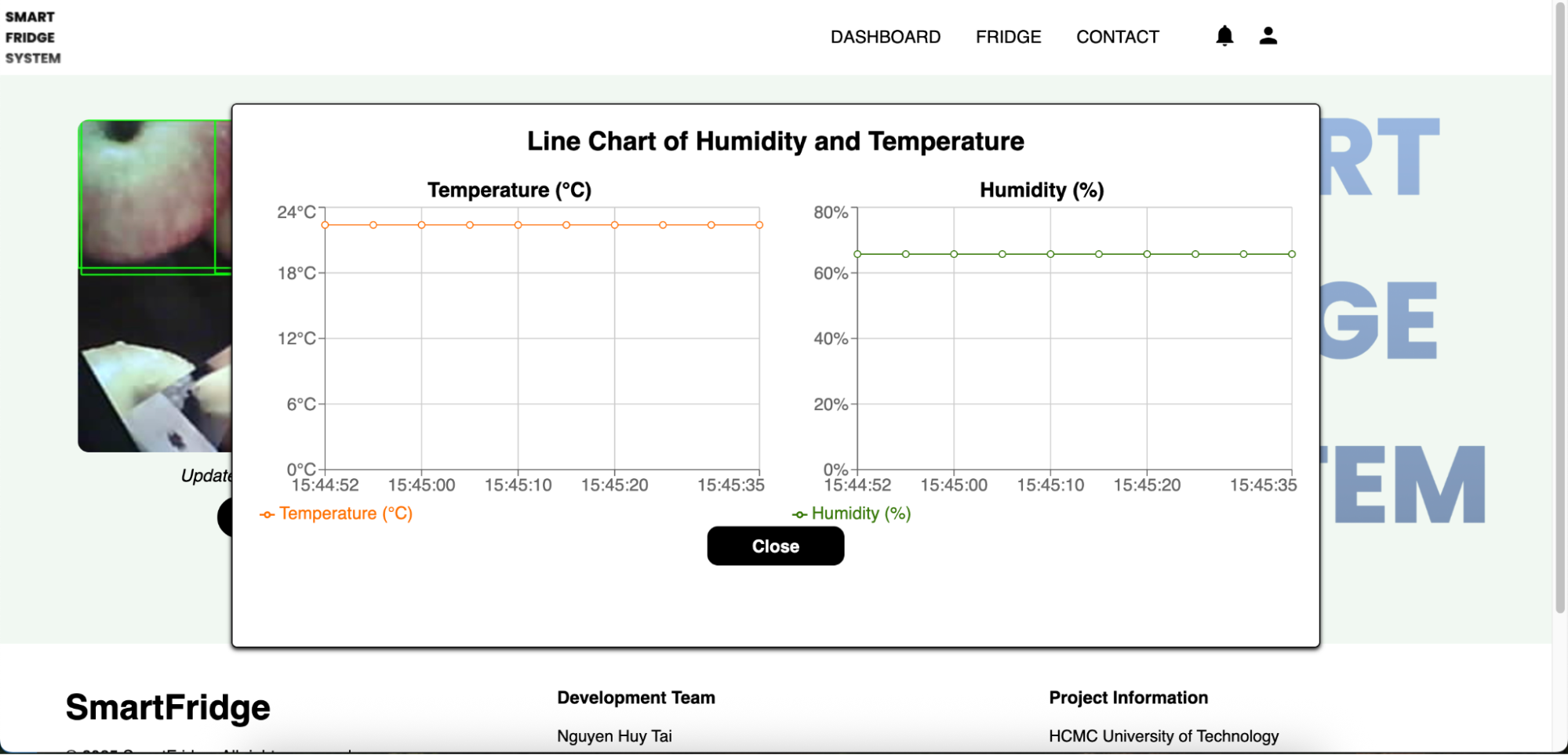}
    \caption{Graphs of temperature and humidity inside the fridge}
    \label{fig:graphs_web}
\end{figure}

The system effectively visualizes both temperature and humidity inside the refrigerator. These values are not only displayed in real time but are also recorded and presented through interactive line charts, which show in \textbf{\autoref{fig:graphs_web}}, enabling users to monitor environmental changes over time and adjust the temperature and humidity levels as desired to ensure optimal storage conditions.

\section{Conclusions}
This study presents the deployment of an AI-driven system for managing food inventory in a refrigerator through a web-based platform. To enhance the model's calibration and performance, we integrated and evaluated three loss functions: Binary Cross-Entropy (BCE), Calibration-aware Focal Loss (CALFOCAL), and Adaptive Focal Loss (ADAFOCAL). CALFOCAL and ADAFOCAL are variants of the traditional Focal Loss, designed to improve model calibration and adaptability during training.

Experimental results reveal that BCE achieves better calibration by producing outputs closely aligned with the ideal diagonal in the reliability diagram, indicating well-calibrated confidence scores. In contrast, both CALFOCAL and ADAFOCAL demonstrate underconfidence, where the predicted probabilities are lower than the actual accuracies, despite their design to handle class imbalance and calibration.

These findings suggest that while adaptive focal loss functions offer advantages in certain contexts, they may require further tuning or modification to achieve optimal confidence calibration in food recognition tasks. Future work will explore the development or selection of loss functions that strike a balance between improving confidence calibration and maintaining robustness, aiming to make the model more confident without becoming overconfident. Additionally, expanding the dataset and incorporating uncertainty estimation techniques could further improve model reliability in real-world applications.

\section*{Acknowledgments}
This research is funded by Office for International Study Programs (OISP), Ho Chi Minh City University of Technology (HCMUT), VNUHCM under grant number .... We acknowledge Ho Chi Minh City University of Technology (HCMUT), VNUHCM for supporting this study.

%Bibliography
\bibliographystyle{unsrt}  
\bibliography{references}  

\end{document}